\documentstyle[11pt]{article}
\title{Square root singularity in the viscosity of neutral colloidal
suspensions at large frequencies}
\author{R. Verberg and I. M. de Schepper\\
I. R. I. Delft University of Technology\\
2629 JB Delft, The Netherlands\\
and\\
M. J. Feigenbaum and E. G. D. Cohen\\
The Rockefeller University\\
New York, NY 10021}
\date{  }
\begin{document}
\maketitle

\begin{abstract}

The asymptotic frequency, $\omega$, dependence of the dynamic viscosity
of neutral hard sphere colloidal suspensions is shown to be of the form 
$\eta_0 A(\phi) (\omega \tau_P)^{-1/2}$, where $A(\phi)$ has been determined 
as a function of the volume fraction $\phi$, for all concentrations 
in the fluid range, $\eta_0$ is the solvent viscosity and $\tau_P$ 
the P\'{e}clet time.
For a soft potential it is shown that, to leading order in the
steepness, the asymptotic behavior is
the same as that for the hard sphere potential and a condition for the
cross-over behavior to $1/\omega \tau_P$ is given. 
Our result for the hard sphere potential generalizes a result of Cichocki and 
Felderhof obtained at low 
concentrations$^{\cite{cich1991}}$ and agrees
well with the experiments of van der Werff et al$^{\cite{werf1989}}$,
if the usual Stokes-Einstein diffusion coefficient $D_0$
in the Smoluchowski operator is consistently replaced by the short-time 
self diffusion coefficient $D_s(\phi)$ for non-dilute colloidal
suspensions.
\end{abstract}

\vspace{.5in}

\newpage
\baselineskip=1.5\baselineskip
\newcounter{chanum}
\newcounter{eqnnum1}[chanum]
\renewcommand{\theequation}{\arabic{chanum}.\arabic{eqnnum1}}

\section{Introduction}
\stepcounter{chanum}

The visco-elastic behavior, i.e. the frequency dependent 
viscosity $\eta(\phi,\omega)$, of concentrated neutral hard sphere colloidal
suspensions has been obtained in the benchmark experiments of Van der Werff
et al$^{\cite{werf1989}}$ for volume fractions $0.44 < \phi < 0.58$, where 
$\phi = \pi n \sigma^3/6$, with $n$ the number density of the colloidal 
particles of diameter $\sigma$. The frequency
dependence was found to be qualitatively similar to that obtained
theoretically by Cichocki and Felderhof$^{\cite{cich1991}}$ for dilute 
suspensions
from an exact solution of the two particle Smoluchowski equation for two
Brownian particles without hydrodynamic interactions.

An approximate theory for concentrated colloidal suspensions was developed
by Verberg et al$^{\cite{verbpre}}$ which agreed well with the 
experimental results
of Van der Werff et al for such suspensions.  In particular, the asymptotic
behavior of the (complex) viscosity for large frequencies $\omega$ 
was given
correctly as $\sim \eta_0 A(\phi)(1+i)/\sqrt{\omega \tau_P}$, 
where $\eta_0$ is the viscosity of the solvent, $A(\phi)$ 
an amplitude and $\tau_P$ a characteristic Brownian particle interaction
time, the P\'{e}clet time, defined below.  However, the amplitude $A(\phi)$ 
was at small $\phi$ a factor two smaller than the exact
value obtained by Cichocki and Felderhof at low densities and 
it was too high when 
compared with the experiments of Van der Werff et al at high densities.  This
difference in asymptotic behavior did not affect the good agreement with
experiments carried out in the reduced form used by Van der Werff et 
al$^{\cite{verbpre,cohedavo,sche1993}}$.

In the theory of Verberg et al, $\eta(\phi,\omega)$ was obtained as a sum
of two terms: a short time -- infinite frequency -- contribution 
$\eta_{\infty} (\phi)$ on the very short Brownian-time scale $\tau_B
(\sim 10^{-9}$s) where the Brownian particle forgets its initial velocity 
and a long time contribution, on the very much longer P\'{e}clet-time scale 
$\tau_P (\sim 10^{-4}$s), involving mode-mode coupling contributions associated
with two cage-diffusion modes, that 
describe the diffusion of each colloidal particle out of the cage in which
it finds itself in a concentrated colloidal suspension$^{\cite{verbpre}}$:
\addtocounter{eqnnum1}{1}
\begin{equation}
\eta(\phi,\omega)=\eta_\infty (\phi)+\eta_{mc}(\phi,\omega)
\end{equation}  
For large $\omega$, the mode-mode coupling contributions 
$\eta_{mc}(\phi,\omega)$ reduces to:
\addtocounter{eqnnum1}{1}
\begin{equation}
\eta_{mc}(\phi,\omega)=\frac{9}{5} \phi^2 \chi(\phi)^{5/2}
\frac{1}{\sqrt{\omega \tau_P}} (1+i) \eta_0 + O(\frac{1}{\omega})
\end{equation}
where $\chi(\phi)$ is the equilibrium radial distribution function
$g_{eq}(r;\phi)$ at contact, i.e. $\chi(\phi) =
g_{eq}(r \! =\! \sigma;\phi)$, where $r$ is the distance between two hard 
spheres of diameter $\sigma$ and $\tau_P = \sigma^2/4 D_0$.
Here $D_0$ is the Stokes-Einstein diffusion coefficient
\addtocounter{eqnnum1}{1}
\begin{equation}
D_0 = \frac{k_B T}{3 \pi \eta_0 \sigma}
\end{equation}
where $k_B$ is Boltzmann's constant and $T$ the temperature of the colloidal
suspension.

For low concentrations $\phi \rightarrow 0, \chi(\phi) \rightarrow 1$ and
$\eta_{mc}(\phi,\omega)$ reduces to:
\addtocounter{eqnnum1}{1}
\begin{equation}
\eta_{mc}(\phi,\omega)=\eta(\phi,\omega) - \eta_{\infty}(\phi)
\stackrel{\phi \rightarrow 0}{\longrightarrow}\frac{9}{5}
\frac{\phi^2}{\sqrt{\omega \tau_P}}(1+i) \eta_0
\end{equation}   
while Cichocki and Felderhof obtain$^{\cite{cich1991}}$:
\addtocounter{eqnnum1}{1}
\begin{equation}
\eta_{CF}(\phi,\omega) -\eta_\infty(\phi) 
\stackrel{\phi \rightarrow 0}{\longrightarrow} \frac{18}{5}
\frac{\phi^2}{\sqrt{\omega \tau_P}}(1+i) \eta_o
\end{equation}
The different coefficient in eq.(1.4) for the
approach to $\eta_{\infty}(\phi)$ is
due to the approximate nature of $\eta_{mc}(\phi,\omega)^{\cite{verbpre}}$.

The purpose of this paper
is to obtain the exact asymptotic behavior of $\eta (\phi,\omega)$ for
large $\omega$ for {\it{all}} $\phi$ studied by van der Werff et 
al$^{\cite{werf1989}}$, i.e. an extension of Cichocki and 
Felderhof's result to high concentrations, as well as its
behavior for a soft potential.

In the next section we will give the basic equations. In section 3
we will calculate the asymptotic frequency dependent viscosity, for
a soft, but very steep potential, starting
from the Green-Kubo expression. In section 4 we will give the result for
a hard sphere potential, as the limit of a soft potential. We will 
end with a short discussion on the soft potential result.

\section{Basic equations}
\stepcounter{chanum}

In order to obtain the asymptotic behavior of $\eta (\phi,\omega)$ for
large $\omega$ for concentrated suspensions we start from a general 
Green-Kubo expression for the frequency dependent viscosity of a 
colloidal suspension$^{\cite{feld1987}}$:
\addtocounter{eqnnum1}{1}
\begin{equation}
\eta(\phi,\omega)=\eta_{\infty}(\phi) + \frac{\beta}{V} \int_0^\infty dt
\rho_\eta(t;\phi) e^{i\omega t}
\end{equation}
Eq.(2.1) gives the linear response of the suspension to an applied shear rate
$\gamma(t) = \gamma_0 e^{-i\omega t}$ with finite frequency $\omega$ and
vanishing shear rate amplitude $\gamma_0 \rightarrow 0$.
In eq.(2.1), $\beta = 1/k_B T$, 
$V$ is the volume of the colloidal suspension, 
while $\rho_{\eta}(t;\phi)$ is the stress-stress
auto correlation function defined by:
\addtocounter{eqnnum1}{1}
\begin{equation}
\rho_\eta(t;\phi) = <\Sigma_{xy}^{\eta} (r^N)e^{t \Omega (r^N;\phi)} 
\Sigma^\eta_{xy} (r^N) >_{eq}
\end{equation}
where the brackets denote a canonical equilibrium ensemble average. 
The microscopic 
stress tensor $\Sigma_{xy}^{\eta}(r^N)$ is given by:
\addtocounter{eqnnum1}{1}
\begin{equation}
\Sigma_{xy}^{\eta}(r^N) = \sum_{i=1}^{N} r_{i,x}F_{i,y}(r^N)
\end{equation}
with ${\bf r}_i$ the position of particle $i \; (i=1, ..., N), \; 
r^N={\bf{r}}_1,..., {\bf{r}}_N, \; 
{\bf F}_i = -\nabla_i\Phi(r^N)$ the total force on particle $i$ ($\nabla_i =
\partial/\partial{\bf r_i}$), $\Phi(r^N)$ the total potential energy of the 
colloidal particles and:
\addtocounter{eqnnum1}{1}
\begin{equation}
\Omega (r^N;\phi) = \sum_{i,j=1}^{N} (\nabla_i + \beta {\bf F}_i(r^N)) 
\cdot {\bf D}_{ij}(r^N) \cdot \nabla_j
\end{equation}
the $N$-particle Smoluchowski 
operator$^{\cite{puse1985,puse1991,russ1989}}$, 
the colloidal analogue 
of the Liouville operator for atomic liquids$^{\cite{hans1986}}$, with 
${\bf D}_{ij}(r^N)$ the diffusion tensor, incorperating hydrodynamic 
interactions. This 
diffusion tensor determines the velocity imparted to 
particle $i$ by a force acting on particle $j$. In the absence of 
hydrodynamic interactions, i.e. for $\phi \rightarrow 0$, the diffusion tensor 
becomes diagonal and independent of $r^N$,
\addtocounter{eqnnum1}{1}
\begin{equation}
{\bf D}_{ij}(r^N) = D_0 {\bf 1} \delta_{ij}
\end{equation}
with {\bf 1} the unit tensor and $\delta_{ij}$ the Kronecker symbol.
However, in concentrated suspensions, where hydrodynamic interactions no longer
can be neglected, ${\bf D}_{ij}(r^N)$ becomes a function of  
the positions of all particles, involving therefore many-particle interactions.

The diffusion tensor ${\bf D}_{ij}(r^N)$ is directly related to 
the experimental short time self-diffusion coefficient 
$D_s(\phi)$ by$^{\cite{puse1985,puse1991}}$
\addtocounter{eqnnum1}{1}
\begin{equation}
D_s(\phi) \equiv < {\bf \hat{Q}} \cdot {\bf D}_{ii}(r^N) \cdot 
{\bf \hat{Q}} >_{eq}
\end{equation}
for any particle $i$. Here ${\bf \hat{Q}}$ is a 3-dimensional 
unit vector. $D_s(\phi)$
reduces in the dilute limit to the Stokes-Einstein diffusion coefficient
$D_0$ (cf.eq.(2.5)). $D_s(\phi)$ is a purely hydrodynamic quantity, which 
involves the calculation of the very complicated many-particle interactions, 
and has been the subject of research for many years, both 
theoretically$^{\cite{cich1988} - \cite{cler1992}}$ and 
experimentally$^{\cite{kops1981} - \cite{otte1987}}$. By now the behavior of 
${\bf D}_{ij}(r^N)$ and $D_s(\phi)$ for intermediate volume fractions up 
to $\phi \approx 0.45$, is fairly well understood theoretically. For higher 
concentrations to the best of our knowledge only
semi-empirical results exist.

However, in this paper we are particularly interested in the high volume 
fractions $\phi > 0.40$. Therefore we were forced to incorporate hydrodynamic 
interactions in an approximate (mean-field-like) fashion, using eq.(2.6). 
This approximation seems justified for high frequencies, where the 
particle distribution in the
suspension is very close to the equilibrium particle distribution, so 
that the hydrodynamic interactions are described in first order by the
hydrodynamic interactions of the suspension at infinite frequency.
The mean-field approximation
will be done explicitly in the next section. For $D_s(\phi)$ we use at the end 
a semi-empirical relation which is consistent with experiments and
hard sphere computer simulations for all 
$\phi$ up to $\phi \approx 0.60^{\cite{brad1993}}$.

\section {Large frequency viscosity for soft potentials}
\stepcounter{chanum}

We will proceed with the calculation of the asymptotic behavior of 
$\eta(\phi,\omega)$ for large $\omega$ by calculating the stress-stress
auto correlation function $\rho_{\eta}(t;\phi)$ of eq.(2.2) with the 
microscopic stress tensor of eq.(2.3) and the $N$-particle Smoluchowski
operator of eq.(2.4) for a soft, but very steep potential.

We restrict ourselves to pairwise additive 
potentials, i.e. $\Phi(r^N) = \Sigma_{i<j=1}^{N} V(r_{ij})$, with 
$V(r_{ij})$ the two 
particle potential and ${\bf{r}}_{ij} = {\bf{r}}_i - {\bf{r}}_j$,
$r_{ij} = |{\bf{r}}_{ij}|$. We can then write eqs.(2.3) and (2.4), 
respectively as
\addtocounter{eqnnum1}{1}
\begin{equation}
\Sigma_{xy}^{\eta}(r^N) = -\frac{1}{2} \sum_{i\neq j}^{N} r_{ij,x} 
\frac{\partial V(r_{ij})}
{\partial r_{i,y}}
\end{equation}
and
\addtocounter{eqnnum1}{1}
\begin{equation}
\Omega(r^N;\phi) = \sum_{i,j,k=1}^{N} \left( \delta_{ik} \nabla_i -
\beta (1-\delta_{ik}) \frac{\partial V(r_{ik})}{\partial {\bf{r}}_i}
\right) \cdot {\bf D}_{ij}(r^N) \cdot \nabla_j
\end{equation}

We determine the asymptotic behavior of $\eta(\phi,\omega)$ from 
eq.(2.1) for a soft, but steep pair potential 
$V_{l}(r) = \epsilon (\sigma/r)^{l}$, with $\epsilon$ the pair 
interaction energy at $r = \sigma$ and $l = |r \frac{\partial}
{\partial r} \ln V_{l}(r) |$ the steepness of the potential.
Since the hard sphere potential is the limit for $\ell \rightarrow \infty$ of 
$V_{\ell}(r)$, one can obtain the hard sphere result, by letting
$l \rightarrow \infty$ at the end of the calculation. This is discussed 
in the next section.

In order to compute $\rho_{\eta}(t;\phi)$ for short time,
we first expand $\rho_{\eta}(t;\phi)$ for $V_l(r)$ 
in powers of $t$. Thus, we write first eq.(2.2)
for $\rho_{\eta}(t;\phi)$ as
\addtocounter{eqnnum1}{1}
\begin{equation}
\rho_{\eta}(t;\phi) = \sum_{n=0}^{\infty} \frac{1}{n!} t^n \rho_{\eta}^{n}(\phi)
\end{equation}
and then calculate $\rho_{\eta}^{n}(\phi) = < \Sigma_{xy}^{\eta}(r^N) \Omega^n
(r^N;\phi) \Sigma_{xy}^{\eta}(r^N) >_{eq}$, using eq.(3.1) for the 
second $\Sigma_{xy}^{\eta}(r^N)$ within the brackets, i.e.:
\addtocounter{eqnnum1}{1}
\begin{equation}
\rho_{\eta}^{n}(\phi) = -\frac{1}{2}N(N-1) < \Sigma_{xy}^{\eta} (r^N) 
\Omega^n(r^N;\phi) 
r_{12,x} \frac{\partial V_l(r_{12})}{\partial r_{1,y}} >_{eq}
\end{equation}
for the soft potential $V_l(r)$.

In this paper we will restrict ourselves to only the leading order 
in the steepness. Thus, we take into account, for each 
$n$, only the most divergent terms in $l$. This implies that, for
short times, we can neglect all the 
contributions of more than two particles to the equilibrium ensemble average 
$< \: >_{eq}$, i.e., we only have to calculate the two-particle 
contributions, involving the equilibrium radial distribution function 
$g_{eq}(r;\phi)$. This, because, relative to the two particle contributions,
the $n$-particle contributions are of the order $l^{-(n-2)}$, so that they can
be neglected in the limit $l \rightarrow \infty^{\cite{sche1981}}$.

Thus, for $\Omega(r^N;\phi)$
in eq.(3.4) we can use eq.(3.2) with only $i,j \in \{1,2\}$, giving
\addtocounter{eqnnum1}{1}
\begin{eqnarray}
\lefteqn{ \rho_{\eta}^{n}(\phi)= -\frac{1}{2}N(N-1) }
\nonumber \\
 & \cdot < \Sigma_{xy}^\eta (r^N)
\left[ \left( \nabla_1 - \beta \frac{\partial V_l(r_{12})} {\partial {\bf r}_1} 
\right) \cdot {\bf D}_r(r^N) \cdot \nabla_1 \right]^n r_{12,x}
\frac{\partial V_l(r_{12})}{\partial r_{1,y}} >_{eq}
\end{eqnarray}
Here we used the symmetry of $\Omega(r^N;\phi)$ in the 
particles 1 and 2, when applied on functions of ${\bf{r}}_{12}$. We have
introduced the relative diffusion coefficient of two interacting spheres
${\bf D}_r(r^N) = 2({\bf D}_{11}(r^N) - {\bf D}_{12}
(r^N))^{\cite{russ1989,brad1993}}$. In the dilute
limit for just two particles, distant from all others, the diffusion tensors
${\bf D}_{11}({\bf r}_1,{\bf r}_2)$ and ${\bf D}_{12}({\bf r}_1,{\bf r}_2)$ 
are known$^{\cite{cich1988,feld1977,jeff1984}}$.

For concentrated suspensions we make a mean-field approximation. We replace 
${\bf D}_r(r^N)$ in eq.(3.5) by its mean value $<{\bf D}_r(r^N)>_{eq}$,
which reduces for high frequencies, i.e., for short times to twice the 
single particle short time self-diffusion 
coefficient $D_s(\phi)$ as given in 
eq.(2.6)$^{\cite{brad1993,cich1994,brad1994}}$. Thus we write in eq.(3.5),
\addtocounter{eqnnum1}{1}
\begin{eqnarray}
{\bf D}_r(r^N) = <{\bf D}_r(r^N)>_{eq} = 2D_s(\phi){\bf 1}
\end{eqnarray}
consistently with eq.(2.6).

Using this approximation and eq.(3.1) for $i,j \in \{1,2\}$, i.e. neglecting 
again all but two-particle contributions, we find straight forwardly:
\addtocounter{eqnnum1}{1}
\begin{eqnarray}
\lefteqn{ \rho_{\eta}^{n}(\phi) = \frac{1}{2}N(N-1) (2D_s(\phi))^n } \nonumber \\
 & & \cdot < r_{12,x} \frac{\partial V_l(r_{12})}
{\partial r_{1,y}} \left[ \left( \nabla_1 - \beta \frac{\partial V_l(r_{12})}
{\partial {\bf r}_1} \right) \cdot \nabla_1 \right]^n r_{12,x}
\frac{\partial V_l(r_{12})}{\partial r_{1,y}} >_{eq}
\end{eqnarray}
Since $\nabla_1 {\bf{r}}_{12} V_l(r_{12}) = {\bf{r}}_{12} 
\nabla_1 V_l(r_{12})(1+O(l^{-1}))$, we can shift, to leading order in
the steepness,
the differential operator $\nabla_1$ through ${\bf r}_{12}$ in any product 
of ${\bf r}_{12}$ and $V_l(r_{12})$ or its derivatives. Thus, from eq.(3.7)
we obtain:
\addtocounter{eqnnum1}{1}
\begin{eqnarray}
\lefteqn{\rho_{\eta}^{n}(\phi) = \frac{1}{2}N(N-1) (2D_s(\phi))^n } 
\nonumber \\
& & \cdot < \frac{r_{x}^{2}r_{y}^{2}}{r^2}
V_{l}'(r)  \left[ \left( \nabla_1 - \beta V_{l}'(r) {\bf \hat{r}} \right)
\cdot \nabla_1 \right]^n V_{l}'(r) >_{eq}
\end{eqnarray}
with ${\bf r} = {\bf r}_{12}$, $r = |{\bf r}_{12}|$, ${\bf \hat{r}} = 
{\bf r}/r$ ,
${\partial V_l(r_{12})}/{\partial r_{12}} = V_{l}'(r)$. 

Changing to spherical coordinates, using the definition of the equilibrium 
radial distribution function $g_{eq}(r;\phi)$ for the soft potential
$V_l(r)$ and performing the angular 
integration we find from eqs.(3.3) and (3.7):
\addtocounter{eqnnum1}{1}
\begin{equation}
\rho_{\eta}(t;\phi) = \frac{2}{15} \pi n^2V \int_{0}^{\infty} dr 
g_{eq}(r;\phi) r^4V_{l}'(r) e^{2 t D_s (\phi)(\nabla_{r}^{2} - 
\beta V_{l}'(r) \nabla_r)} V_{l}'(r)
\end{equation}
where $\nabla_r = \partial/\partial r$.

Thus we have expressed $\rho_{\eta}(t;\phi)$ at large volume fractions 
$\phi$ and to leading order in the steepness $l$ in terms of a 
one dimensional integral over $r$ involving 
the high density equilibrium radial distribution function 
$g_{eq}(r;\phi)$ for a potential $V_l(r)$ of finite but large $l$ 
and the effective short time self-diffusion
coefficient $D_s(\phi)$. In the next section we consider the hard sphere
limit $(l \rightarrow \infty)$ of eq.(3.9).

\section {Hard sphere limit}
\stepcounter{chanum}

To evaluate eq.(3.9) we introduce the function
\addtocounter{eqnnum1}{1}
\begin{equation}
y_{eq}(r;\phi) = g_{eq}(r;\phi) e^{\beta V_l (r)}
\end{equation}
As discussed in ref.{\cite{hans1986}}, $y_{eq}(r;\phi)$ is,
unlike $g_{eq}(r;\phi)$, a smooth continuous function of $r$ 
for {\it{all}} $\; r$ and $l$. For hard spheres $g_{eq}^{hs}(r;\phi) =
y_{eq}^{hs}(r;\phi)$ for $r \geq \sigma$ and $\chi(\phi) \equiv 
g_{eq}^{hs}(r=\sigma;\phi) = y_{eq}^{hs}(r=\sigma;\phi)$ is the pair
correlation function at contact.
Writing $V_{hs}(r) =  V_{l \rightarrow \infty}(r)$, using that
\addtocounter{eqnnum1}{1}
\begin{equation}
e^{- \beta V_{hs}(r)}V_{hs}'(r) = -\frac{1}{\beta} \delta(r-\sigma)
\end{equation}
and using eq.(4.1), we can write for the stress-stress auto correlation 
$\rho_{\eta}(t;\phi)$ of eq.(3.9) in the hard-sphere limit 
$l \rightarrow \infty$:
\addtocounter{eqnnum1}{1}
\begin{equation}
\rho_{\eta}(t;\phi) = \frac{2}{15} \pi n^2 V \chi(\phi) \int_{0}^{\infty} 
dr r^4 e^{-\beta V_{hs}(r)} V_{hs}'(r) e^{\Omega^{hs}_{r} t} V_{hs}'(r)
\end{equation}
Here we have defined $\Omega^{hs}_{r} = 2 D_s(\phi) (\nabla_{r}^{2} - 
\beta V_{hs}'(r) \nabla_r)$, the radial part of the two-particle 
Smolochowski operator for a hard-sphere potential in relative coordinates.

This expression can be calculated by a method, similar to that of 
Cichocki and Felderhof$^{\cite{cich1991}}$. The actual calculation 
is given in more detail in the Appendix, giving:
\addtocounter{eqnnum1}{1}
\begin{equation}
\rho_{\eta}(t;\phi) = \frac{18}{5} \phi^2 \chi(\phi) \left( \frac{2 D_0}
{\pi D_s(\phi)} \right)^{\frac{1}{2}} \frac{V}{\beta}  
\frac{1}{\sqrt{t \tau_P}} \eta_0
\end{equation}
Eq.(4.4) leads, with eq.(2.1), to our final result for hard spheres: 
\addtocounter{eqnnum1}{1}
\begin{equation}
\eta(\phi,\omega) \stackrel{\omega \rightarrow \infty}{\longrightarrow}
\eta_{\infty}(\phi) + A(\phi) \frac{1+i}{\sqrt{\omega \tau_P}} \eta_0 
\end{equation}
with the coefficient of the square root singularity $A(\phi)$ given by:
\addtocounter{eqnnum1}{1}
\begin{equation}
A(\phi) = \frac{18}{5} \phi^2 \chi(\phi) \left( \frac{D_0}{D_s(\phi)} 
\right)^{\frac{1}{2}}
\end{equation}
This result, based on the Green-Kubo formula for the frequency dependent 
viscosity $\eta(\phi,\omega)^{\cite{feld1987}}$ of a colloidal suspension 
consisting of 
hard spheres with hydrodynamic interactions included, is compared 
with experiments of van der Werff et al$^{\cite{werf1989}}$ in fig.1.
Here we have used for $\chi(\phi)$ the Carnahan-Starling
approximation$^{\cite{brad1993,hansp3695}}$ $\chi(\phi) = 
(1-0.5\phi)/(1-\phi)^3$ for 
$\phi \leq 0.5$ and a one-pole approximation$^{\cite{brad1993}}$ 
$\chi(\phi) = 1.2/(1-\phi / \phi_m)$ for $\phi > 0.5$, with $\phi_m = 0.63$
the volume fraction at random close packing.
For $D_s(\phi)/D_0$ we have used Beenakker and Mazur's
expression$^{\cite{been1984}}$
for $\phi \leq 0.45$ and $D_s(\phi)/D_0 = 0.85(1-\phi / \phi_m)$
for $\phi > 0.45^{\cite{brad1993}}$.

Figure 1 clearly shows that in order to obtain 
agreement with experiment it is neccesary to include hydrodynamic 
interactions, i.e., to take into account the diffusion tensor 
${\bf D}_{ij}(r^N)$, in the basic N-particle Smoluchowski equation.
In a mean-field approximation this leads to the replacement of
the Stokes-Einstein diffusion coefficient $D_0$ by the short time 
self-diffusion coefficient $D_s(\phi)$, a replacement also 
made by Brady$^{\cite{brad1993}}$.
Eqs.(4.5) and (4.6) reduce to the exact expression (eq.(1.5)) obtained 
before by Cichocki and Felderhof for low densities$^{\cite{cich1991}}$
to $O(\phi^2)$ (see Appendix).

We will show in the next section (eqs.(5.1) and (5.3)) that the
right hand side of eq.(4.4) is the leading term 
of the expansion in powers of $l$ of $\rho(t;\phi)$ for a soft
potential, i.e., for finite $l$ and for frequencies $\omega$ up
to $\sim \frac{l^2}{\tau_P} \frac{D_s(\phi)}{D_0}$.

\section{Soft potential}
\stepcounter{chanum}

We note that the large $\omega$-behavior of $\eta(\phi,\omega)
\sim 1/\sqrt{\omega}$ is typical for hard spheres. For any soft, but
steep potential $\eta(\phi,\omega) \sim 1/\omega$ for 
$\omega \rightarrow \infty$. 
For example, the presence of a lubrication layer causes a change in the 
relative diffusion of two spheres at very short times, which leads
to $1/\omega$ behavior at very high frequencies, as discussed by
Cichocki and Felderhof$^{\cite{cich1994a}}$ and Rallison and 
Hinch$^{\cite{rall1986}}$.
To study the transition from the $1/\omega$-behavior (for 
any finite $l$) to the $1/\sqrt{\omega}$-behavior ($l = \infty$) we have
calculated $\rho_\eta(t;\phi)$ of eq.(3.9) for finite $l$.
For finite $l$ eq.(3.9) can no longer be calculated by a method,
similar to that of Cichocki and Felderhof$^{\cite{cich1991}}$,
described in the appendix, but involves the calculation of the complete 
eigenvalue problem of the radial part of $\Omega_r = 2D_s(\phi)(\nabla_r^2 -
\beta V_l'(r)\nabla_r)$, the two-particle Smoluchowski
operator in relative coordinates, for finite $l^{\cite{verbfeig}}$.
The result for $\rho_\eta(t;\phi)$ in eq.(3.9) can be written 
as$^{\cite{verbfeig}}$
\addtocounter{eqnnum1}{1}
\begin{equation}
\rho_{\eta}(t;\phi) = \frac{2 \pi n^2 V l \sigma^3 \chi(\phi)}{15 \beta^2}
r(\tau(\phi))
\end{equation}
with $\tau(\phi) = 2D_s(\phi) tl^2/ \sigma^2$. The function $r(x)$ can
be expanded for $x \ll 1$ as:
\addtocounter{eqnnum1}{1}
\begin{equation}
r(x) = 1 - x + \frac{3}{2} x^2 + O(x^3)
\end{equation}
and for $x > 1$ as$^{\cite{verbfeig}}$:
\addtocounter{eqnnum1}{1}
\begin{equation}
r(x) = \frac{1}{\sqrt{\pi x}} \left( 1 + \frac{\pi^2}{12x} + 
O(\frac{1}{x^2}) \right)
\end{equation}
Thus to leading order in the steepness $l$, we obtain from eqs.(5.1)
and (5.3):
\addtocounter{eqnnum1}{1}
\begin{equation}
\rho_{\eta}(t;\phi) = \frac{2 \pi n^2 V l \sigma^3 \chi(\phi)}{15 \beta^2}
\frac{1}{\sqrt{\pi \tau(\phi)}}, \; \; \; \; \; 
t > \frac{\tau_P}{l^2} \frac{D_0}{D_s(\phi)}
\end{equation}
consistent with eq.(4.4) when $l \rightarrow \infty$. For finite $l$,
$\eta(\phi,\omega) \sim 1/ \sqrt{\omega}$ for
frequencies $\omega$ up to $\sim \frac{l^2}{\tau_P}
\frac{D_s(\phi)}{D_0}$, while for larger $\omega$, 
$\eta(\phi,\omega)$ behaves as
$1/\omega$, as is typical for soft potentials.
It might be interesting to see whether such a transition in the
asymptotic behavior of $\eta(\phi,\omega)$ can be observed in 
concentrated colloidal
suspensions, where the interaction potential is steep.

\vspace{.3in}
{\large{\bf{Acknowledgement}}} 
E. G. D. C. acknowledges financial support under grant DE-FG02-88-ER13847
from the U.S. Department of Energy and R. V. support from the Netherlands 
Foundation for Fundamental Research of Matter (FOM).

\newpage
\noindent{\bf{\Large{Appendix}}}

\vspace{.3in}
\newcounter{eqnappx}
\renewcommand{\theequation}{A.\arabic{eqnappx}}
\setcounter{eqnappx}{1}

Here we calculate the stress-stress autocorrelation function
$\rho_{\eta}(t;\phi)$ for hard spheres.
We start with eq.(4.3) for $\rho_{\eta}(t;\phi)$ in
the hard-sphere limit, i.e. $l \rightarrow \infty$:
\addtocounter{eqnappx}{1}
\begin{equation}
\rho_{\eta}(t;\phi) = \frac{2}{15} \pi n^2V \chi(\phi) \int_{0}^{\infty} 
dr r^4 e^{-\beta V_{hs}(r)} V_{hs}'(r) e^{\Omega^{hs}_{r} t} V_{hs}'(r)
\end{equation}
with $V_{hs}(r)$ the hard-sphere potential and
\addtocounter{eqnappx}{1}
\begin{equation}
\Omega^{hs}_r = 2D_s(\phi)(\nabla_{r}^{2} - \beta V_{hs}'(r) \nabla_r)
\end{equation}
Using that $\exp(- \beta V_{hs}(r)) V_{hs}'(r) = -\frac{1}{\beta} 
\delta(r-\sigma)$ gives
\addtocounter{eqnappx}{1}
\begin{equation}
\rho_{\eta}(t;\phi) = - \frac{2 \pi n^2 V \sigma^4 \chi(\phi)}{15 \beta} 
\int_{0}^{\infty} dr \delta(r-\sigma) e^{\Omega^{hs}_{r} t} V_{hs}'(r)
\end{equation}
With eq.(2.1), eq.(A.4) gives
\addtocounter{eqnappx}{1}
\begin{eqnarray}
\eta(\phi,\omega) &=& \eta_\infty(\phi) - \frac{2}{15} \pi n^2 \sigma^4 
\chi(\phi) \int_{0}^{\infty} dt \int_{0}^{\infty} dr \delta(r-\sigma) 
e^{(\Omega^{hs}_{r} +i\omega) t} V_{hs}'(r) \nonumber \\
&=& \eta_\infty + \frac{2}{15} \pi n^2 \sigma^4 \chi(\phi) 
\int_{0}^{\infty} dr \delta(r-\sigma) \frac{1}{(\Omega^{hs}_{r} 
+i\omega)} V_{hs}'(r)
\end{eqnarray}
We define
\addtocounter{eqnappx}{1}
\begin{equation}
f(r,\omega) = \frac{1}{(\Omega^{hs}_{r} +i\omega)} V_{hs}'(r)
\end{equation}
and deduce the following differential equation for $f(r,\omega)$:
\addtocounter{eqnappx}{1}
\begin{eqnarray}
\left( 2D_s \nabla_{r}^{2} - 2D_s \beta V_{hs}'(r) \nabla_r + i \omega \right)
f(r,\omega) = V_{hs}'(r)
\end{eqnarray}
Due to the singular behavior of the hard-sphere potential at $r= \sigma$,
eq.(A.7) reduces to the boundary value problem:
\addtocounter{eqnappx}{1}
\begin{equation}
\left\{ \begin{array}{lc}
(2D_s \nabla_{r}^{2} + i \omega)f(r,\omega) = 0 \; \; \; & r > \sigma \\
2D_s \beta \nabla_r f(r,\omega) = -1 \; \; \; & r = \sigma
\end{array} \right.
\end{equation}
with the solution, bounded for $r \geq \sigma$
\addtocounter{eqnappx}{1}
\begin{equation}
f(r,\omega) = \frac{1}{2 \alpha \beta D_s} e^{-\alpha(r-\sigma)}
\end{equation}
where $\alpha^2 = -i \omega / 2D_s$. Eq.(A.5) with eqs.(A.6) and (A.9) 
gives the asymptotic result for the frequency dependent viscosity as
given in eqs.(4.4) and (4.5)
\addtocounter{eqnappx}{1}
\begin{equation}
\eta(\phi,\omega) = \eta_{\infty}(\phi) + \frac{18}{5} \phi^2 \chi(\phi) 
\left( \frac{D_0} {D_s(\phi)} \right)^{\frac{1}{2}} 
\frac{1+i}{\sqrt{\omega \tau_P}} \eta_0
\end{equation}
where we have used the P\'{e}clet time $\tau_P = \sigma^2 /4D_0$ and the
Stokes-Einstein relation for $D_0$ as given in eq.(1.3).
Eq.(A.10) with eq.(2.1) yields for $\rho_\eta(t;\phi)$:
\addtocounter{eqnappx}{1}
\begin{equation}
\rho_{\eta}(t;\phi) = \frac{18}{5} \phi^2 \chi(\phi) \left( \frac{2 D_0}
{\pi D_s(\phi)} \right)^{\frac{1}{2}} \frac{V}{\beta}  
\frac{1}{\sqrt{t \tau_P}} \eta_0
\end{equation}
For low concentrations ($\phi \rightarrow 0$), $\chi(\phi) = 1$ and
$D_s(\phi)= D_0$ and $\rho_{\eta}(t;\phi)$ reduces to the result of Cichocki
and Felderhof$^{\cite{cich1991}}$ for $\rho_{\eta}(t;\phi \rightarrow 0)$, 
for short times. These authors calculate $\rho_{\eta}(t;\phi \rightarrow 0)$
for hard spheres on the basis of eq.(2.2) restricted from the 
beginning to two hard-sphere particles only,
but for all times $t$. One can show that the angular dependences in
eq.(2.2) (i.e. in $\Sigma_{xy}$) are irrelevant for short times.
Thus, for $t \rightarrow 0$, both approaches are similar, leading 
to identical results for $\rho_{\eta}(t;\phi)$ and $\eta(\phi,\omega)$.

\newpage
{\large{\bf{Figure captions}}}\\

\noindent {\underline{Figure 1.}} Coefficient $A(\phi)$ of the square root 
singularity in $\omega$, i.e., $\sim A(\phi) (1+i)\eta_0 / \sqrt{\omega 
\tau_P}$, of $\eta(\phi,\omega)$ as a function of the volume 
fraction $\phi$. Experimental points $(\bullet)$ of van der Werff et 
al, ref.{\cite{werf1989}}; The vertical lines indicate the estimated
errors in the experimental values of $A(\phi)$, while the horizontal 
lines indicate the effect of the 4\%
uncertainty in $\phi$ (ref.{\cite{kruipriv}}). The dashed line represents the 
mode-coupling result eq.(1.2) and the solid line our result given in
eq.(4.6). Here we have used for $\chi(\phi)$ the Carnahan-Starling
approximation$^{\cite{hansp3695}}$ $\chi(\phi) = (1-0.5\phi)/(1-\phi)^3$ for 
$\phi \leq 0.5$ and a one-pole approximation$^{\cite{brad1993}}$ 
$\chi(\phi) = 1.2/(1-\phi / \phi_m)$ for $\phi > 0.5$, with $\phi_m = 0.63$
the volume fraction at random close packing.
For $D_s(\phi)/D_0$ we have used Beenakker and Mazur's
expression$^{\cite{been1984}}$
for $\phi \leq 0.45$ and $D_s(\phi)/D_0 = 0.85(1-\phi / \phi_m)$
for $\phi > 0.45^{\cite{brad1993}}$.
The dotted line represents eq.(4.6) with $D_s(\phi) = D_0$, i.e.,
when hydrodynamic interactions are neglected.

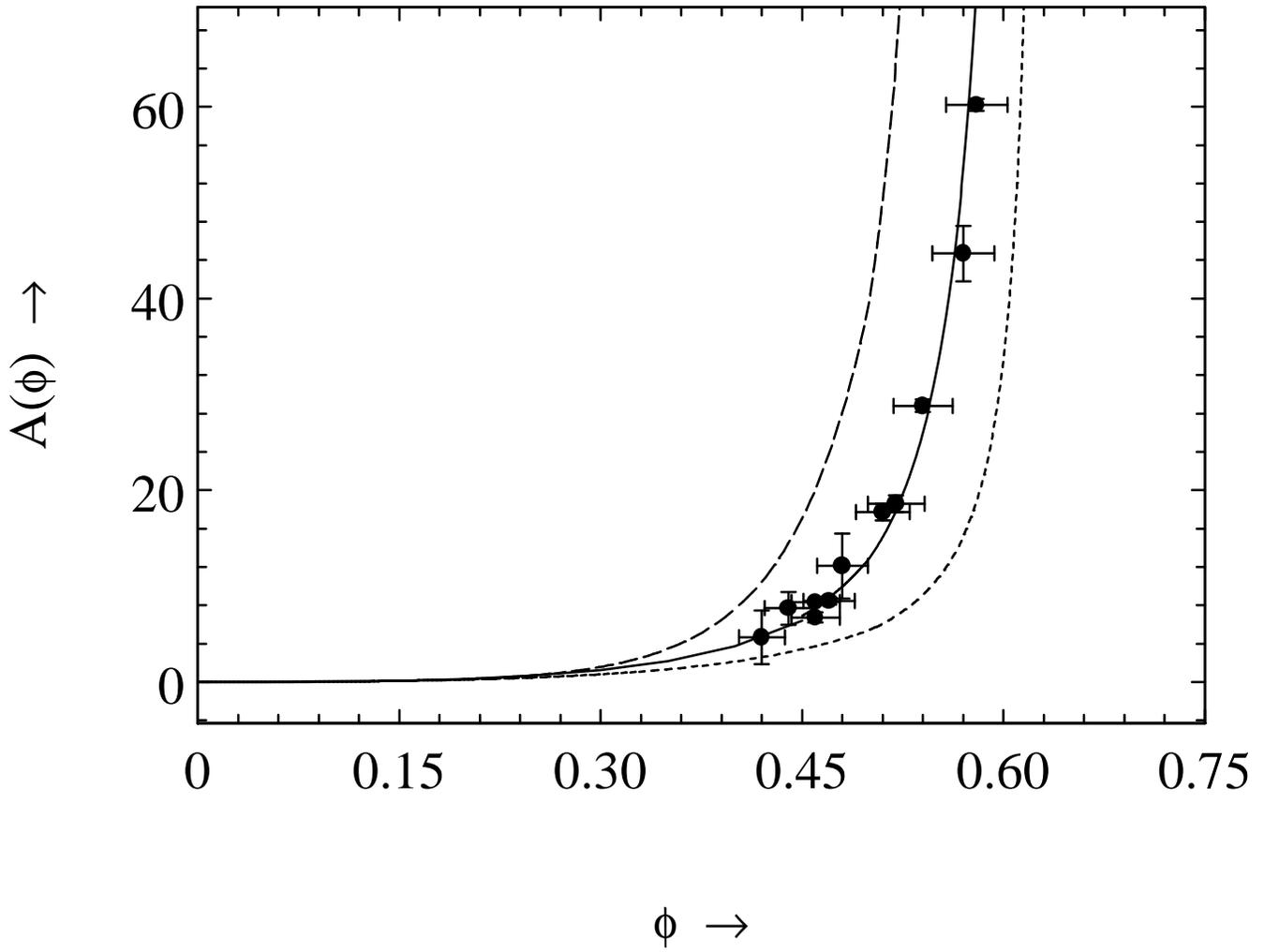
\end{document}